\documentclass[runningheads,a4paper]{llncs}
\usepackage[utf8]{inputenc}

\usepackage{amssymb}
\usepackage{graphicx}
\usepackage{subfig}
\usepackage{multirow}
\usepackage{textcomp} 
\usepackage{cite}

\usepackage{url}
\urldef{\mailsa}\path|{chgferreira, brenomatos, jussara}@dcc.ufmg.br|    
\newcommand{\keywords}[1]{\par\addvspace\baselineskip
\noindent\keywordname\enspace\ignorespaces#1}

\begin{document}

\mainmatter  

\title{
Analyzing Dynamic Ideological Communities in  Congressional Voting Networks}

\titlerunning{Analyzing Ideological Communities in Congressional Voting Networks}

%
%
\author{Carlos H. G. Ferreira\footnote{\url{https://orcid.org/0000-0001-9107-6884}}%
\and Breno de Sousa Matos\and Jussara M. Almeira}
\authorrunning{Analyzing Ideological Communities in Congressional Voting Networks}

\institute{Departament of Computer Science, Universidade Federal de Minas Minas Gerais\\ Belo Horizonte - Brazil\\
\mailsa}

%
%

\toctitle{Analyzing Ideological Communities in Congressional Voting Networks}

\tocauthor{Carlos H. G. Ferreira, Breno de Souza Matos and Jusssara M. Almeida}

\maketitle
\vspace{-0.2cm}

\begin{abstract}
We here study the behavior of political party members aiming at identifying how ideological communities are created and evolve over time in diverse (fragmented and non-fragmented) party systems.  Using public voting data of both Brazil and the US, we propose a methodology to identify and characterize ideological communities, their member polarization,  and how such communities evolve over time, covering a 15-year period.  Our results reveal very distinct patterns across the two case studies, in terms of both structural  and  dynamic properties. 

\end{abstract}
\section{Introduction}


Party systems can be characterized based on their fragmentation and polarization  \cite{Sartori:2005}. Party fragmentation corresponds to the number of parties existing in a political system (e.g., a country) while polarization is related to the multiple opinions that lead to the division of members into groups with distinct political ideologies  \cite{Sartori:2005, Morris:2008}. In countries where the party system has a low fragmentation, the polarization of political parties can be seen more clearly since one party tends to occupy more seats supporting the government and the other opposes it \cite{Mann:2017}. On the other hand, in fragmented systems the multiple political parties often make use of coalitions, a type of inter-party alliance, to raise their relevance in the political system and reach a common end \cite{Ames:2009, Budge:2016}. Thus, a great amount of ideological similarity, as expressed by their voting decisions, is often observed across different parties.

Previous work has analyzed the behavior of political party  members through the modeling of voting data in signed and weighted networks \cite{Andris:2015, Porter:2005, Moody:2013, Dal:2015, Cherepnalkoski:2016, Arinik:2017, Levorato:2017, Mendonca:2017}. These prior efforts tackled topics such as community detection, party cohesion and loyalty analysis, governance of a political party and member influence in such networks. However, the identification and characterization of ideological communities, particularly in fragmented party systems, require observing some issues, such as: (i) presidents define coalitions throughout government in order to strengthen the implementation of desired public policies, which may be ruptured after a period of time \cite{Mainwaring2:1997, Budge:2016}; (ii) political members have different levels of partisanship and loyalty, and their political preferences may change over time \cite{Baldassarri:2008, Andris:2015}; and (iii) different political parties may have the same political ideology, being redundant under a party system \cite{Vaz:2015}.


In such context, we here study the behavior of political party members aiming at identifying how ideological communities are created and evolve over time. To that end, we consider two case studies, Brazil and US, which are representatives of distinct party systems: whereas the former is highly fragmented and redundant \cite{Vaz:2015}, the latter is not fragmented but rather polarized with two major parties, although some party members can be considered less polarized \cite{Moody:2013, Dal:2015}. Using public datasets of the voting records in the House of Representatives of both countries during a 15-year period, we  characterize the emergence and evolution of communities of House members with similar political ideology (captured by their voting behavior)  by using complex network concepts. Specifically, we tackle three research questions (RQs):

\begin{itemize}
    \item \textbf{RQ1: How do ideological communities are characterized in governments with different (i.e., fragmented and non-fragmented) party systems?}  
    We model the voting behavior of each House of Representatives during a given time period using a  network, where nodes represent House members, and weighted edges are added if two members voted similarly. We use the Louvain algorithm \cite{Blondel:2008} to detect communities in each network and characterize  structural properties of such communities. Unlike prior community analysis in the political context, we compare the properties of these communities in fragmented and in non-fragmented party systems.
    
    
     \item \textbf{RQ2: How can we identify polarization in the ideological communities?} 
 We use neighborhood overlap \cite{Easley:2010} to estimate the tie strength associated to each network edge, characterizing it as either strong or weak. This approach to estimate tie strength has been employed in several contexts \cite{Granovetter:1977, Marlow:2013, McGee:2013, Wiese:2015} and also in the political context \cite{Waugh:2009}. However, these prior studies were not interested in analyzing and comparing distinct political systems, as we do here. We use strong ties to identify polarized communities in each analyzed network.


 \item \textbf{RQ3: How do polarized communities evolve over time?}
  We analyze how polarized communities evolve over the years of a government, characterizing how the membership of such communities change over time.
\end{itemize}

In sum, the key contributions of our work are: (i) a methodology to identify and analyze dynamic ideological communities and their polarization in party systems based on complex network concepts; and (ii) two case studies covering strikingly different party systems over a quite broad time period. Our study shows that in fragmented party systems, such as Brazil, although party redundancy exists, some ideological communities exist and may, indeed, be polarized. However, such polarized communities are highly dynamic, greatly changing their membership over consecutive years.  In the US, on the other hand, despite the strong and  temporally stable party polarization, there are members, within each party, that exhibit different levels of polarization.


The rest of this paper is organized as follows. Section 2 briefly discusses related work, whereas Section 3 describes our modeling methodology and case studies. We then present our main results, tackling RQ1-RQ3 in Sections 4-6. Conclusions and future work are offered in Section 7.

\section{Related Work}

Complex networks constitute a set of theoretical and analytical tools to describe and analyze phenomena related to interactions  occurring in the real world \cite{Rossetti:2018}. Among the many properties of a network, the interactions between  pairs of nodes can be used to define the strength of these links (or {\it tie strength}) \cite{Easley:2010}.
Indeed, tie strength is a property that has been widely studied in several domains.
For example,  the tie strength between pairs of people was studied in the phone call and Short Message System (SMS) networks, where a higher frequency of SMS and longer call duration characterize stronger ties \cite{Wiese:2015}.  The different types of interactions between Facebook users have also been used to define tie strength on that system \cite{Marlow:2013}. 
Similarly, tie strength was used to build geolocation models based on Twitter data  and exploited in the prediction of user location \cite{McGee:2013}. 


In the political context, the study of political ideologies has been largely accomplished through the analysis of roll call votes networks. In a roll call votes network, the nodes represent people (e.g., congressmen), and two nodes are connected if they have voted similarly in one (or more) voting sessions.  In \cite{Andris:2015}, the authors studied the committee's formation in the US House of Representatives using roll call votes networks, finding that there is a cooperation between the Democratic and Republican parties. Although the polarization in recent decades has been increasing, there are moderate members in both parties, who cooperate with each other. In the same direction, authors in \cite{Porter:2005} studied the committees and subcommittees of US House of Representatives exploiting the network connections that are built according to common membership. Analogously, the polarization in the US Senate was evaluated using a network defined by the similarity of Senators' votes \cite{Moody:2013}. 

In \cite{Dal:2015}, the authors studied the relations between members  of the Italian parliament according to their voting behavior, analyzing the community structure with respect to political coalitions and government alliances over time. Similarly, the cohesiveness of members of the European Parliament was investigated through the analysis of network models combining roll call votes and Twitter data \cite{Cherepnalkoski:2016}. Other approaches study the behavior of political members modeling roll call votes using signed networks. However, this type of analysis is appropriate for modeling only polarized systems \cite{Arinik:2017}. Signed networks have also been used to evaluate aspects related to political governance and political party behavior \cite{Levorato:2017}. In addition, an algorithm was proposed to evaluate signed networks and a case study was conducted using a European Parliament network capturing voting similarities between members \cite{Mendonca:2017}. In \cite{Levorato:2017}, the roll call votes of  the Brazilian House of Representative was modeled using signed networks. The results revealed inefficient coalitions with the government as parties that make such coalitions  have members distributed in different ideological communities over time. 
Orthogonally, others have investigated the ideology of political members and users through profiles of social networks \cite{Agathangelou:2017, Darwish:2017, Wang:2017}. 

Unlike prior work, our focus here is on the characterization of ideological  communities   in {\it diverse}, i.e.,  both fragmented and non-fragmented, party systems. We also propose to use tie strength, computed based on neighborhood overlap, to identify polarized communities under the party systems diversities, evaluating their evolution over time, on both party systems.


\section{Methodology}
In this section, we describe the methodology used in our study. We start by presenting basic concepts (Section \ref{subsection:basic}) followed by our case studies (Section \ref{subsection:cases}), and then describe our modeling of voting behavior (Section \ref{subsection:modeling}).

\subsection{Basic Concepts}\label{subsection:basic}

The House of Representatives is composed of several members who occupy the seats during each government period. House members participate in a series of voting sessions, when bills, amendments, and propositions are discussed and voted. Thus, attending such sessions is the most direct way for members  to express their ideologies and positioning. When these members are associated with a large number of political parties, the party system in question is regarded as fragmented. In this case, during a term of office, coalition governments are established, leading political parties to organize themselves into ideological communities, defending together common interests during voting sessions \cite{Mainwaring2:1997, Sartori:2005}.

One can evaluate the behavior of parties and their members  in terms of how cohesive they are as an ideological community by analyzing voting data using widely disseminated metrics, such as  Rice's Index. However, the use of Rice Index has been shown to be problematic when there are more than two voting options (other than only {\it yes} and {\it no}) \cite{Hix:2005}, as is the case, for example, in the European Parliament and in our study, as we will see later.   

Instead, we here employ the  {\it Partisan Discipline} and {\it Party Discipline} metrics \cite{Vaz:2015}. The former captures the ideological alignment of a   member to her party (estimated by the behavior of the majority), and the latter expresses the  ideological cohesiveness of a party. Given a member $m$, belonging to party $p_m$, the {\it Partisan Discipline} of $m$, $pd_m$ is given by the fraction of all voting sessions to which $m$ attended and voted similarly to the majority of $p_m$'s members. That is, let $n$ be the number of voting sessions attended by member $m$ and  $I(m,p_m,i)$ be 1 if member $m$ voted similarly to the majority of members of $p_m$ in voting session $i$ ($i=1..n$) and 0 otherwise. Then:

\vspace{-0.2cm}
\begin{equation}
    pd_m = \frac{\sum_{i=1}^{n} I(m,p_m,i)}{n}
\end{equation}
\vspace{-0.2cm}

The \textit{Partisan Discipline} can be generalized to assess the discipline and ideological alignment of a  member to any community (not only his original party). 

The {\it Party Discipline} of a party $p$ is computed as the average {\it Partisan Discipline} of all of its members, that is, $PD(p) = \frac{\sum_{m=1}^{M} pd_m}{M}$, where $M$ is the number of members of $p$.  {\it Party Discipline} captures how cohesive a party (or community) is in a set of votes. Both metrics range from 0 to 1, where 1 indicates that a member or party is totally disciplined (or cohesive) and 0 otherwise. 
\subsection{Case Studies}
\label{subsection:cases}

We consider two case studies: Brazil and US.
In Brazil, the House of Representatives consists of 513 seats. A member vote can be either \textit{Yes}, \textit{No}, \textit{Obstruction} or \textit{Absence} in each voting session. A \textit{Yes} or \textit{No} vote expresses, respectively, an agreement or disagreement with the given proposition. Both \textit{Absence} and {\it Obstruction} mean that the member did not participate in the voting, although  an \textit{Obstruction} expresses the intention of the member to cause the voting session to be cancelled due to insufficient quorum. Similarly, the US House of Representatives includes 435 seats, and a member vote  can be \textit{Yes}, \textit{No} or \textit{Not Voting}, whereas the last one indicates the member was not present in the voting session. In our study, we disregard {\it Absence} and {\it Not Voting} votes, as they do not reflect any particular inclination of the members with respect to the topic under consideration. However, we do include {\it Obstructions}  as they reflect an intentional action of the  members and a clear opposition to the topic. Thus, for Brazil, three different voting options were considered.

For both case studies, we collected voting data from public sources. 
The plenary roll call votes of Brazil's House of Representatives are available through an application programming interface (API) maintained by the government\footnote{\url{http://www2.camara.leg.br/transparencia/dados-abertos/dados-abertos-legislativo} (In Portuguese)}. We collected roll call votes between the $52^{th}$ and $55^{th}$ legislatures, from 2003 to 2017. US voting data covering the same 15-year period (i.e., between the $108^{th}$ and $115^{th}$ congresses) was collected through the  ProPublica API\footnote{\url{https://projects.propublica.org/api-docs/congress-api/}}.
Each dataset consists of a sequence of the voting session; for each session,  the dataset includes  date, time and voting option of each participating member.

\begin{table}[t]
\centering
\scriptsize
\caption{Overview of our datasets.}
\label{tab:Vote_Data}
\resizebox{.9\columnwidth}{!}{
\begin{tabular}{|c|c|c|c|c|c|c|c|c|}
\hline
\multicolumn{9}{|c|}{\textbf{Brazil}} \\ \hline                                                  
\textbf{Leg.}              & \textbf{Year} & \textbf{\begin{tabular}[c]{@{}c@{}}President\\ (Party)\end{tabular}} & \textbf{\begin{tabular}[c]{@{}c@{}}\# of Voting\\ Sessions\end{tabular}} & \textbf{\begin{tabular}[c]{@{}c@{}}\# of \\ Votes\end{tabular}} & \textbf{\begin{tabular}[c]{@{}c@{}}\# of \\ Parties\end{tabular}} & \textbf{\begin{tabular}[c]{@{}c@{}}\# of \\ Members\end{tabular}} & \textbf{\begin{tabular}[c]{@{}c@{}}Avg.\\ PD\end{tabular}} & \textbf{\begin{tabular}[c]{@{}c@{}}SD\\ PD\end{tabular}} \\ \hline
\multirow{4}{*}{$52^{th}$} & 2003          & Lula (PT)                                                            & 150                                                                      & 106755                                                          & 23                                                                & 435                                                               & 88.23\%                                                      & 0.08                                                     \\ \cline{2-9} 
                           & 2004          & Lula (PT)                                                            & 118                                                                      & 71576                                                           & 23                                                                & 377                                                               & 87.43\%                                                      & 0.08                                                     \\ \cline{2-9} 
                           & 2005          & Lula (PT)                                                            & 81                                                                       & 50616                                                           & 24                                                                & 382                                                               & 88.91\%                                                      & 0.07                                                     \\ \cline{2-9} 
                           & 2006          & Lula (PT)                                                            & 87                                                                       & 62358                                                           & 24                                                                & 419                                                               & 91.12\%                                                      & 0.05                                                     \\ \hline
\multirow{4}{*}{$53^{th}$} & 2007          & Lula (PT)                                                            & 221                                                                      & 190424                                                          & 31                                                                & 478                                                               & 92.45\%                                                      & 0.07                                                     \\ \cline{2-9} 
                           & 2008          & Lula (PT)                                                            & 157                                                                      & 122482                                                          & 31                                                                & 452                                                               & 92.34\%                                                      & 0.07                                                     \\ \cline{2-9} 
                           & 2009          & Lula (PT)                                                            & 156                                                                      & 125759                                                          & 30                                                                & 465                                                               & 91.87\%                                                      & 0.06                                                     \\ \cline{2-9} 
                           & 2010          & Lula (PT)                                                            & 83                                                                       & 63255                                                           & 29                                                                & 452                                                               & 92.46\%                                                      & 0.05                                                     \\ \hline
\multirow{4}{*}{$54^{th}$} & 2011          & Dilma (PT)                                                           & 98                                                                       & 78662                                                           & 29                                                                & 481                                                               & 89.34\%                                                      & 0.08                                                     \\ \cline{2-9} 
                           & 2012          & Dilma (PT)                                                           & 79                                                                       & 60219                                                           & 28                                                                & 454                                                               & 89.56\%                                                      & 0.05                                                     \\ \cline{2-9} 
                           & 2013          & Dilma (PT)                                                           & 158                                                                      & 115751                                                          & 29                                                                & 451                                                               & 88.70\%                                                      & 0.06                                                     \\ \cline{2-9} 
                           & 2014          & Dilma (PT)                                                           & 87                                                                       & 66154                                                           & 28                                                                & 451                                                               & 92.93\%                                                      & 0.04                                                     \\ \hline
\multirow{3}{*}{$55^{th}$} & 2015          & Dilma (PT)                                                           & 273                                                                      & 231031                                                          & 28                                                                & 502                                                               & 85.84\%                                                      & 0.06                                                     \\ \cline{2-9} 
                           & 2016          & \begin{tabular}[c]{@{}c@{}}Dilma (PT)\\ Temer (PMDB)\end{tabular}    & 218                                                                      & 156006                                                          & 28                                                                & 452                                                               & 90.12\%                                                      & 0.05                                                     \\ \cline{2-9} 
                           & 2017          & Temer (PMDB)                                                         & 230                                                                      & 159704                                                          & 29                                                                & 435                                                               & 89.76\%                                                      & 0.08                                                     \\ \hline \hline \hline

\multicolumn{9}{|c|}{\textbf{United States}}                                                                                      \\ \hline
\textbf{Cong.}              & \textbf{Year} & \textbf{\begin{tabular}[c]{@{}c@{}}President\\ (Party)\end{tabular}} & \textbf{\begin{tabular}[c]{@{}c@{}}\# of Voting\\ Sessions\end{tabular}} & \textbf{\begin{tabular}[c]{@{}c@{}}\# of \\ Votes\end{tabular}} & \textbf{\begin{tabular}[c]{@{}c@{}}\# of \\ Parties\end{tabular}} & \textbf{\begin{tabular}[c]{@{}c@{}}\# of \\ Members\end{tabular}} & \textbf{\begin{tabular}[c]{@{}c@{}}Avg.\\ PD\end{tabular}} & \textbf{\begin{tabular}[c]{@{}c@{}}SD\\ PD\end{tabular}} \\ \hline
\multirow{2}{*}{$108^{th}$} & 2003          & George W. Bush (R)                                                   & 623                                                                      & 258867                                                          & 3                                                                 & 432                                                               & 95.76\%                                                      & 0.03                                                     \\ \cline{2-9} 
                            & 2004          & George W. Bush (R)                                                   & 502                                                                      & 203557                                                          & 3                                                                 & 427                                                               & 95.11\%                                                      & 0.03                                                     \\ \hline
\multirow{2}{*}{$109^{th}$} & 2005          & George W. Bush (R)                                                   & 637                                                                      & 264735                                                          & 3                                                                 & 432                                                               & 95.02\%                                                      & 0.03                                                     \\ \cline{2-9} 
                            & 2006          & George W. Bush (R)                                                   & 511                                                                      & 210592                                                          & 3                                                                 & 428                                                               & 94.98\%                                                      & 0.04                                                     \\ \hline
\multirow{2}{*}{$110^{th}$} & 2007          & George W. Bush (R)                                                   & 956                                                                      & 297957                                                          & 2                                                                 & 414                                                               & 92.23\%                                                      & 0.04                                                     \\ \cline{2-9} 
                            & 2008          & George W. Bush (R)                                                   & 605                                                                      & 244734                                                          & 2                                                                 & 426                                                               & 92.73\%                                                      & 0.04                                                     \\ \hline
\multirow{2}{*}{$111^{th}$} & 2009          & Barack Obama (D)                                                     & 929                                                                      & 385344                                                          & 3                                                                 & 431                                                               & 93.78\%                                                      & 0.02                                                     \\ \cline{2-9} 
                            & 2010          & Barack Obama (D)                                                     & 631                                                                      & 253296                                                          & 3                                                                 & 422                                                               & 95.34\%                                                      & 0.01                                                     \\ \hline
\multirow{2}{*}{$112^{th}$} & 2011          & Barack Obama (D)                                                     & 908                                                                      & 377601                                                          & 2                                                                 & 428                                                               & 91.98\%                                                      & 0.01                                                     \\ \cline{2-9} 
                            & 2012          & Barack Obama (D)                                                     & 621                                                                      & 253812                                                          & 2                                                                 & 425                                                               & 91.50\%                                                      & 0.01                                                     \\ \hline
\multirow{2}{*}{$113^{th}$} & 2013          & Barack Obama (D)                                                     & 594                                                                      & 245430                                                          & 2                                                                 & 427                                                               & 93.04\%                                                      & 0.01                                                     \\ \cline{2-9} 
                            & 2014          & Barack Obama (D)                                                     & 531                                                                      & 217822                                                          & 2                                                                 & 426                                                               & 93.24\%                                                      & 0.01                                                     \\ \hline
\multirow{2}{*}{$114^{th}$} & 2015          & Barack Obama (D)                                                     & 662                                                                      & 277732                                                          & 2                                                                 & 432                                                               & 94.87\%                                                      & 0.01                                                     \\ \cline{2-9} 
                            & 2016          & Barack Obama (D)                                                     & 588                                                                      & 241263                                                          & 2                                                                 & 427                                                               & 95.11\%                                                      & 0.01                                                     \\ \hline
$115^{th}$                  & 2017          & Donald Trump (R)                                                     & 708                                                                      & 292503                                                          & 2                                                                 & 427                                                               & 95.99\%                                                      & 0.00                                                     \\ \hline

\end{tabular}}
\vspace{-0.6cm}
\end{table}

In a preliminary analysis of the datasets, we noted that some members had little attendance to the voting sessions, especially in Brazil. Thus, we chose to filter our datasets to remove  members with low attendance as they introduce noise to our analyses. Specifically, we removed members that had not attended (thus had not associated vote) to more than 33\% of the voting sessions during each year\footnote{This threshold was chosen based on Article 55 of the Brazilian Constitution that establishes that a deputy or senator will lose her mandate if she does not attend more than one third of the sessions.}. On average,  19\% and 1.98\% members were removed from the Brazilian and US datasets for each year, respectively.



Table \ref{tab:Vote_Data} shows an overview of both (filtered) datasets, with Brazil on the top part of the table and the US on the bottom. The table presents each year covered, the acting president\footnote{ Brazilian president Dilma Rousseff was impeached from office and, therefore, Brazil had two Presidents that year.} and his/her party\footnote{For Brazil: Worker's Party (PT) and the Brazilian Democratic Movement Party (PMDB). For the US: Democratic (D) and Republican (R).}, total number of voting sessions, total number of member votes, as well as  numbers of parties and members occupying seats in the House of Representatives during the year.  The two rightmost  columns, \textit{Avg. PD} and \textit{SD PD}, present the average and standard deviation of the \textit{Party Discipline} computed across all parties. 

Starting with the Brazilian dataset, we can see that the number of parties occupying  seats  has somewhat grown in recent years, characterizing an increasingly fragmented party system. However, in general, average \textit{PD} values are very high (ranging from 85\% to 92\%), with small variation across parties, indicating that, despite the fragmentation, most party members have high partisan discipline.
Regarding the American dataset, Table \ref{tab:Vote_Data} shows that the number of voting sessions is much larger than in Brazil. This is because the API of the Brazilian House of Representative  provides only data related to votes in plenary, while the US dataset covers all votes. 
Moreover, although the numbers of members are comparable to those in the Brazilian dataset, the number of parties occupying seats in each year is much smaller. Indeed,
 only two parties, namely Republican (R) and Democrat (D),  fill all seats in the House of Representatives since the $112^{th}$ Congress. Thus, unlike the Brazilian case, party fragmentation is not an issue in the US. Moreover, just like in Brazil,  parties have a high party discipline.

\subsection{Network Model}
\label{subsection:modeling}

We model the  dynamics of ideological communities in voting sessions in each country using graphs as follows. We discretize time into non-overlapping windows of fixed duration.  For each time window $w$ analyzed, we create a weighted and undirected graph $G^w(V,A)$ in which $V$=$\{v_1, v_2, ...v_n\}$ is a set of vertices representing House members and each edge ($v_i$, $v_j$)  is weighted by the similarity of voting positions of members $v_i$ and $v_j$. Specifically, the weight of edge ($v_i$, $v_j$) is given by the ratio of the number of sessions in which both members voted similarly to the total number of sessions to which both members attended, during window $w$.  Since in Brazil, government coalitions are usually made every year, we choose one year as the time window  for analyzing community dynamics. 

\begin{figure}[t]
  \centering
  \subfloat[Brazil]{\includegraphics[width=4cm]{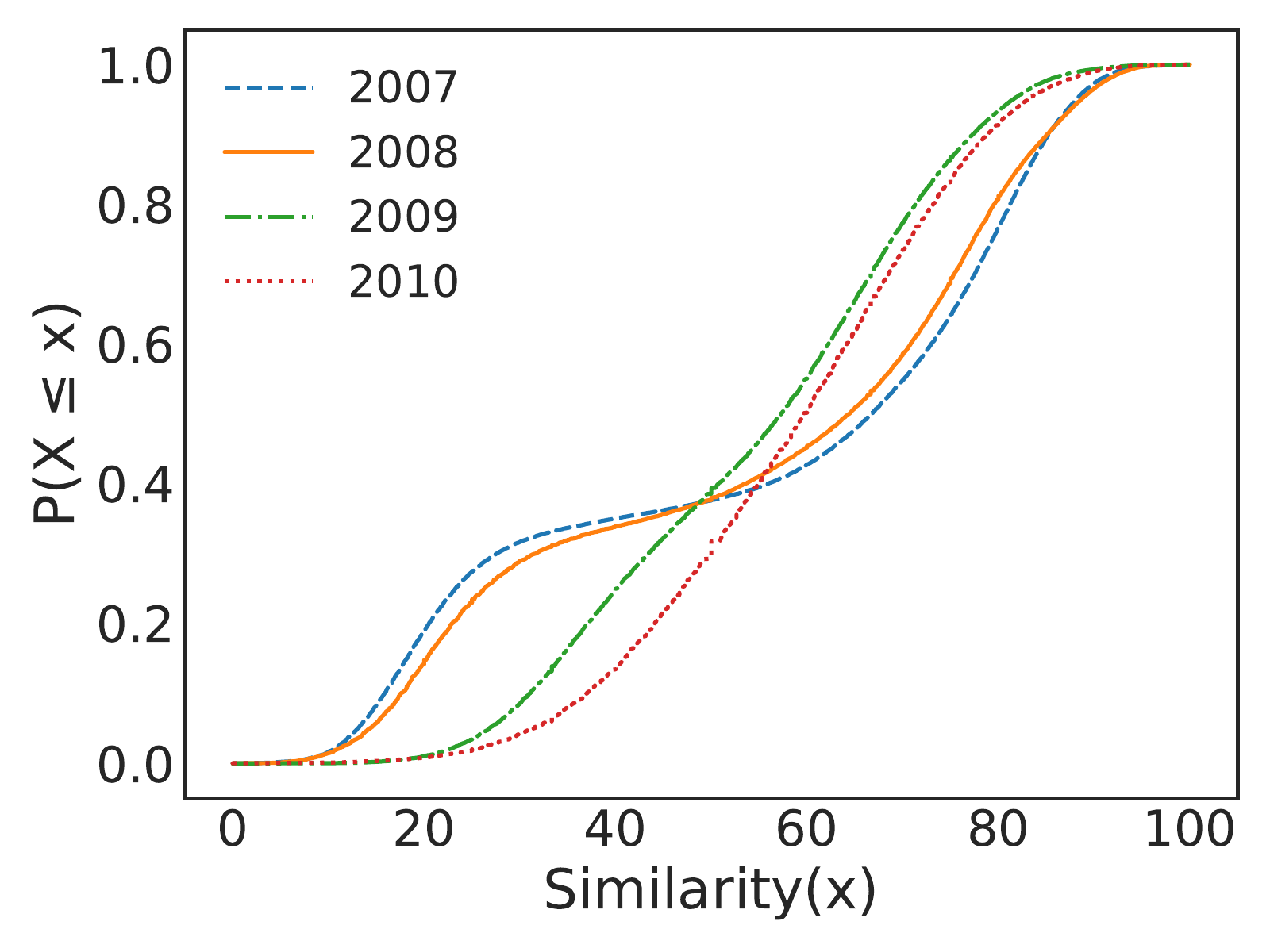} \label{Fig:DIST_BR}}  
  \subfloat[United States]{\includegraphics[width=4cm]{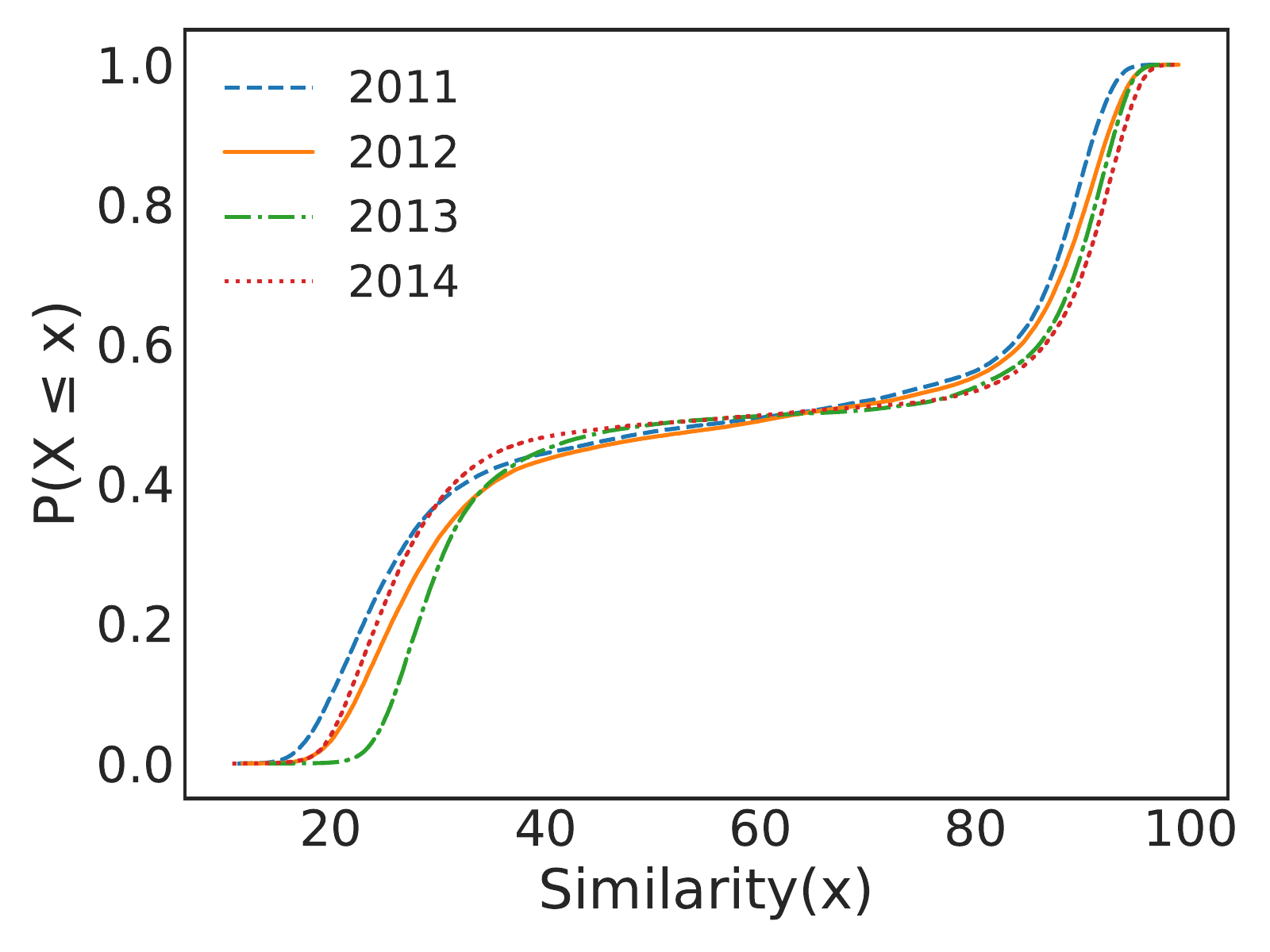} \label{Fig:DIST_US}}
  \vspace{-0.3cm}
  \caption{Cumulative Distribution Function of Edge Similarity.}
\vspace{-0.6cm}
\end{figure}

After building each graph, we noted that all pairs of members voted similarly at least once in all years analyzed and in both countries and, therefore, all graphs built are complete. This reflects the fact that some voting sessions are not discriminative of ideology or opinion, as most members (regardless of party) voted similarly. Thereby, it is necessary to filter out edges that do not contribute to the detection of ideological communities. To that end, we analyzed the distributions of edge similarity for all networks. Representative distributions for specific years, for both Brazil and US, are shown in Figures \ref{Fig:DIST_BR} and \ref{Fig:DIST_US}, respectively.  We note that while the distributions for US exhibit clear concentrations around very small (roughly 0.3) and very large (around 0.85) similarity values, the similarity distributions for Brazil exhibit greater variability, which is consistent with the greater fragmentation of the party system. 

 We argue that, for the sake of removing edges from the graphs,  a similarity threshold should not be much smaller than the average partisan discipline of individual members. That is, two members that have similarity much lower than their partisan disciplines should not be considered as part of the same ideological community.  On the other hand, the higher the similarity threshold chosen, the larger the number of edges removed and the more sparse the resulting graph is. After experimenting with different thresholds, we chose to remove all edges  with  weights  below the $90^{th}$ percentile of the similarity distribution for the Brazilian graphs. For the US,  we removed edges with weights below the $55^{th}$ percentile of the similarity distribution.  Both percentiles correspond roughly to a similarity value of 80\%, which is not much smaller than the average partisan disciplines in both countries (see Table \ref{tab:Vote_Data}). We removed nodes that become isolated after the edge filtering, that is, single-node communities are not included in our analyses.

In sum, we model the voting sessions in each country using two sets of networks, one network per year. Then, we use the Louvain Method  \cite{Blondel:2008} to identify ideological communities in each network. This method has been extensively used to detect network communities in various domains \cite{Santo:2010, Michel:2013, Cai:2016}. It is based on the optimization of \textit{modularity} \cite{Newman:2006}, a metric to evaluate the structure of clusters in a network. \textit{Modularity} is large when the clustering is good and it can reach a maximum value of 1. In this study, we use \textit{modularity} and \textit{party discipline}   as main metrics to assess the cohesiveness of the communities found. The former captures the quality of the result with respect to the topological structure of the communities  in the network, whereas the latter,  computed for the communities (rather than for individual parties), captures quality in terms of context semantics.  In the next sections, we discuss the results of our analyses.

\section{Identifying Ideological Communities}

We start our discussion by tackling our first research question (RQ1) and characterizing the ideological communities discovered in both Brazilian and US networks. Table \ref{tab:RQ1} shows an overview of all networks for both countries, presenting some topological properties \cite{Easley:2010}, i.e., numbers of nodes (\textit{\# of nodes}) and edges (\textit{\# of edges}), number of connected components ({\textit{\# of CC}}), average shortest path length ({\it SPL}), average degree, clustering coefficient  and density\footnote{The \textit{density} of a network is given by the ratio of the total number of existing edges to  the maximum possible number of edges in the graph. The   {\it clustering coefficient}, on the other hand, measures the  degree at which the nodes of the graph tend to group together to form triangles, and is defined as the ratio of the number of existing closed triplets to the total number of open and closed triplets. A triplet is three nodes that are connected by either two (open triplet) or three (closed triplet) undirected ties.}. Note the difference between the number of nodes in this table and the number of members in Table \ref{tab:Vote_Data}, corresponding to nodes that were removed after the edge filtering.

Table \ref{tab:RQ1}  also summarizes the characteristics of the ideological communities identified using the Louvain algorithm. In the four rightmost columns, it presents the number of communities identified, their \textit{modularity} (\textit{Mod.}) as well as average and standard deviation  of the party discipline (\textit{Avg PD} and \textit{SD PD}), computed with respect to the ideological communities. 

Starting with the Brazilian networks (top part of Table \ref{tab:RQ1}), we can observe  great fluctuation in most topological metrics over the years, but, overall, the networks are  sparse: the average shortest path length is short, the average clustering coefficient is moderate and the network density is low. Also, the number of communities identified is much smaller than the total number of parties (see Table \ref{tab:Vote_Data}) confirming the fragmentation and ideological overlap of multiple parties. Yet, the {\it party discipline} of these communities is, on average,  very close to, and, in some cases,  slightly larger than the values computed for the individual  parties, despite a somewhat greater standard deviation observed across communities. Thus, these communities are indeed very cohesive in their voting patterns.

\begin{table}[t]
\centering
\scriptsize
\caption{Characterization of Networks and Ideological Communities}
\label{tab:RQ1}
\resizebox{.9\columnwidth}{!}{
\begin{tabular}{|c|c|c|c|c|c|c|c|c|c|c|c|}
\hline
\multicolumn{12}{|c|}{\textbf{Brazil}}                                                                                      \\ \hline
\textbf{Year} & \textbf{\begin{tabular}[c]{@{}c@{}}\# of \\ Nodes\end{tabular}} & \textbf{\begin{tabular}[c]{@{}c@{}}\# of \\ Edges\end{tabular}} & \textbf{\begin{tabular}[c]{@{}c@{}}\# of \\ CC\end{tabular}} & \textbf{\begin{tabular}[c]{@{}c@{}}Avg.\\ SPL\end{tabular}} & \textbf{\begin{tabular}[c]{@{}c@{}}Avg.\\ Degree\end{tabular}} & \textbf{\begin{tabular}[c]{@{}c@{}}Avg. \\ Clustering\end{tabular}} & \textbf{Density} & \textbf{\begin{tabular}[c]{@{}c@{}}\# of \\ Comm.\end{tabular}} & \textbf{Mod.} & \textbf{\begin{tabular}[c]{@{}c@{}}Avg.\\ PD\end{tabular}} & \textbf{\begin{tabular}[c]{@{}c@{}}SD\\ PD\end{tabular}} \\ \hline

2003          & 342                                                             & 9329                                                            & 5                                                            & 1.83                                                      & 55.01                                                          & 0.65                                                                & 0.16             & 8                                                               & 0.11          & 95.48\%                                                      & 2.22                                                     \\ \hline
2004          & 326                                                             & 7079                                                            & 2                                                            & 1.90                                                      & 43.43                                                          & 0.62                                                                & 0.13             & 4                                                               & 0.14          & 92.68\%                                                      & 3.36                                                     \\ \hline
2005          & 359                                                             & 7211                                                            & 1                                                            & 3.18                                                      & 40.17                                                          & 0.59                                                                & 0.11             & 5                                                               & 0.21          & 88.32\%                                                      & 3.64                                                     \\ \hline
2006          & 419                                                             & 8613                                                            & 1                                                            & 2.47                                                      & 41.11                                                          & 0.61                                                                & 0.09             & 4                                                               & 0.36          & 90.50\%                                                      & 2.36                                                     \\ \hline
2007          & 427                                                             & 11394                                                           & 3                                                            & 1.77                                                      & 53.37                                                          & 0.67                                                                & 0.12             & 6                                                               & 0.14          & 95.97\%                                                      & 1.26                                                     \\ \hline
2008          & 400                                                             & 10180                                                           & 2                                                            & 1.62                                                      & 50.90                                                          & 0.70                                                                & 0.12             & 5                                                               & 0.08          & 95.78\%                                                      & 1.94                                                     \\ \hline
2009          & 434                                                             & 10784                                                           & 2                                                            & 1.92                                                      & 49.70                                                           & 0.66                                                                & 0.11             & 4                                                               & 0.18          & 91.45\%                                                      & 3.49                                                     \\ \hline
2010          & 446                                                             & 10151                                                           & 1                                                            & 2.42                                                      & 45.52                                                          & 0.64                                                                & 0.10             & 4                                                               & 0.19          & 92.01\%                                                      & 1.29                                                     \\ \hline
2011          & 408                                                             & 11519                                                           & 2                                                            & 1.89                                                      & 56.47                                                          & 0.60                                                                & 0.13             & 6                                                               & 0.12          & 93.69\%                                                      & 3.76                                                     \\ \hline
2012          & 345                                                             & 6527                                                            & 3                                                            & 2.47                                                      & 46.11                                                          & 0.48                                                                & 0.11             & 4                                                               & 0.33          & 87.00\%                                                      & 4.25                                                     \\ \hline
2013          & 449                                                             & 10094                                                           & 1                                                            & 2.21                                                      & 44.96                                                          & 0.61                                                                & 0.10             & 4                                                               & 0.38          & 86.51\%                                                      & 4.18                                                     \\ \hline
2014          & 450                                                             & 10036                                                           & 1                                                            & 2.18                                                      & 44.60                                                          & 0.58                                                                & 0.09             & 3                                                               & 0.43          & 91.14\%                                                      & 1.79                                                     \\ \hline
2015          & 490                                                             & 12563                                                           & 1                                                            & 2.90                                                      & 51.28                                                          & 0.69                                                                & 0.10             & 5                                                               & 0.60          & 85.90\%                                                      & 3.11                                                     \\ \hline
2016          & 425                                                             & 10159                                                           & 2                                                            & 1.44                                                      & 47.81                                                          & 0.66                                                                & 0.11             & 4                                                               & 0.38          & 92.62\%                                                      & 1.83                                                     \\ \hline
2017          & 396                                                             & 9434                                                            & 4                                                            & 1.64                                                      & 47.65                                                          & 0.72                                                                & 0.12             & 6                                                               & 0.24          & 90.25\%                                                      & 3.16                                                     \\ \hline \hline \hline
\multicolumn{12}{|c|}{\textbf{United States}}\\ \hline
\textbf{Year} & \textbf{\begin{tabular}[c]{@{}c@{}}\# of \\ Nodes\end{tabular}} & \textbf{\begin{tabular}[c]{@{}c@{}}\# of \\ Edges\end{tabular}} & \textbf{\begin{tabular}[c]{@{}c@{}}\# of \\ CC\end{tabular}} & \textbf{\begin{tabular}[c]{@{}c@{}}Avg.\\ SPL\end{tabular}} & \textbf{\begin{tabular}[c]{@{}c@{}}Avg.\\ Degree\end{tabular}} & \textbf{\begin{tabular}[c]{@{}c@{}}Avg. \\ Clustering\end{tabular}} & \textbf{Density} & \textbf{\begin{tabular}[c]{@{}c@{}}\# of \\ Comm.\end{tabular}} & \textbf{Mod.} & \textbf{\begin{tabular}[c]{@{}c@{}}Avg.\\ PD\end{tabular}} & \textbf{\begin{tabular}[c]{@{}c@{}}SD\\ PD\end{tabular}} \\ \hline
2003          & 431                                                             & 41892                                                           & 2                                                            & 1.11                                                       & 194.39                                                         & 0.95                                                                & 0.45             & 2                                                               & 0.48          & 93.60\%                                                      & 1.03                                                     \\ \hline
2004          & 426                                                             & 40928                                                           & 2                                                            & 1.10                                                       & 192.15                                                         & 0.95                                                                & 0.45             & 2                                                               & 0.48          & 92.97\%                                                      & 0.55                                                     \\ \hline
2005          & 431                                                             & 41892                                                           & 2                                                            & 1.10                                                       & 194.39                                                         & 0.95                                                                & 0.45             & 2                                                               & 0.48          & 92.60\%                                                      & 0.79                                                     \\ \hline
2006          & 426                                                             & 41112                                                           & 2                                                            & 1.10                                                       & 193.01                                                         & 0.95                                                                & 0.45             & 2                                                               & 0.49          & 91.45\%                                                      & 0.33                                                     \\ \hline
2007          & 414                                                             & 38471                                                           & 2                                                            & 1.12                                                       & 185.85                                                         & 0.94                                                                & 0.45             & 2                                                               & 0.44          & 91.55\%                                                      & 3.78                                                     \\ \hline
2008          & 424                                                             & 40729                                                           & 2                                                            & 1.11                                                      & 192.12                                                         & 0.94                                                                & 0.45             & 2                                                               & 0.46          & 95.45\%                                                      & 1.97                                                     \\ \hline
2009          & 429                                                             & 41698                                                           & 2                                                            & 1.15                                                      & 194.40                                                         & 0.94                                                                & 0.45             & 2                                                               & 0.40          & 93.86\%                                                      & 2.42                                                     \\ \hline
2010          & 420                                                             & 39969                                                           & 1                                                            & 3.06                                                       & 190.33                                                         & 0.95                                                                & 0.45             & 3                                                               & 0.43          & 94.92\%                                                      & 1.86                                                     \\ \hline
2011          & 426                                                             & 41119                                                           & 2                                                            & 1.18                                                      & 193.05                                                         & 0.96                                                                & 0.45             & 3                                                               & 0.44          & 90.31\%                                                      & 1.91                                                     \\ \hline
2012          & 417                                                             & 40545                                                           & 3                                                            & 1.17                                                       & 194.46                                                         & 0.96                                                                & 0.46             & 3                                                               & 0.44          & 91.63\%                                                      & 1.86                                                     \\ \hline
2013          & 423                                                             & 40921                                                           & 2                                                            & 1.11                                                       & 193.48                                                         & 0.96                                                                & 0.45             & 2                                                               & 0.47          & 93.23\%                                                      & 1.03                                                     \\ \hline
2014          & 418                                                             & 40735                                                           & 2                                                            & 1.08                                                       & 194.90                                                         & 0.96                                                                & 0.46             & 2                                                               & 0.48          & 94.37\%                                                      & 0.34                                                     \\ \hline
2015          & 427                                                             & 41890                                                           & 2                                                            & 1.09                                                       & 196.21                                                         & 0.95                                                                & 0.46             & 2                                                               & 0.47          & 94.40\%                                                      & 1.36                                                     \\ \hline
2016          & 423                                                             & 40927                                                           & 2                                                            & 1.11                                                      & 193.51                                                         & 0.95                                                                & 0.45             & 2                                                               & 0.48          & 94.70\%                                                      & 1.36                                                     \\ \hline
2017          & 423                                                             & 40928                                                           & 2                                                            & 1.09                                                      & 193.51                                                         & 0.95                                                                & 0.45             & 2                                                               & 0.46          & 96.02\%                                                      & 0.44                                                     \\ \hline
\end{tabular}}
\vspace{-0.6cm}
\end{table}

In contrast,  the topological structure of the identified communities, as expressed by the \textit{modularity} metric, is very weak, especially in the former years. That is, there is still a lot of similarity across members of different ideological communities. We note that in the former years the government had greater  support from most parties, voting similarly in most sessions. Such approval dropped during a period of political turmoil that started in 2012, when the distinction of ideologies and opinions become more clear \cite{Vox:2016, BBC:2018}. This may explain  why the \textit{modularity} starts low and increases in the most recent years, when there is greater distinction between different communities. Note that this happens despite the  large average party discipline maintained by the communities. That is, these two metrics provide complementary interpretations of the political scenario.
 

Turning our attention to the US (bottom part of  Table \ref{tab:RQ1}), we note that, unlike in Brazil, most metrics remain roughly stable throughout the years. The networks are much more dense, with higher average clustering coefficient and density and shortest path length.  The number of identified communities coincides with the number of connected components  as well as with the  number of political parties (see Table \ref{tab:Vote_Data}) in most years. These communities are more strongly structured, despite some ideological overlap, as expressed by moderate-to-large \textit{modularity} value. Moreover, these communities are consistent in their ideologies, as expressed by large party disciplines, comparable to the original (party-level) ones. These metrics reflect the political behavior of a non-fragmented  and stronger two-party system, quite unlike the Brazilian scenario.

In sum,   in Brazil, the several parties can be grouped into just a few ideological communities, with strong disciplined members, although the separation between communities is not very clear. In the US, on the other hand, ideological communities are more clearly defined, both structurally and ideologically, though some inter-community similarity still remains.


\section{Identifying Polarized Communities}

As mentioned, the ideological communities identified in the previous section still share some similarity, particularly for the Brazilian case. 
In this section, we address our second research question (RQ2), with the aim of identifying polarized communities,  i.e., communities that have a more clear distinction from the others in terms of voting behavior. To that end, we take a step further and consider that members of the same polarized community should not only be neighbors (i.e., similar   to each other) but should also share most of their neighbors. Thus two members that, despite voting similar to each other, have mostly distinct sets of neighbors should {\it not} be in the same  group. 

To identify polarized communities, we start with the networks used to identify the ideological communities and compute the {\it neighborhood overlap} for each edge. The neighborhood overlap of an edge ($v_i$, $v_j$) is  the ratio of the  number of nodes that are neighbors of both $v_i$ and $v_j$ to the number of neighbors of at least one of $v_i$ or $v_j$ \cite{Easley:2010}. The neighborhood overlap of $v_i$ and $v_j$ is taken as an estimate of the strength of the tie between the two nodes. Edges with tie strength below (above) a given threshold are classified as {\it weak} ({\it strong}) ties.  We consider that weak ties come from overlapping  communities, and strong ties are edges within a polarized community. Thus, edges representing weak ties are removed. As before, all nodes that become isolated after this second filtering are also removed.

\begin{figure}[t]
  \centering
  \subfloat[\textit{Modularity}]{\includegraphics[width=4.8cm]{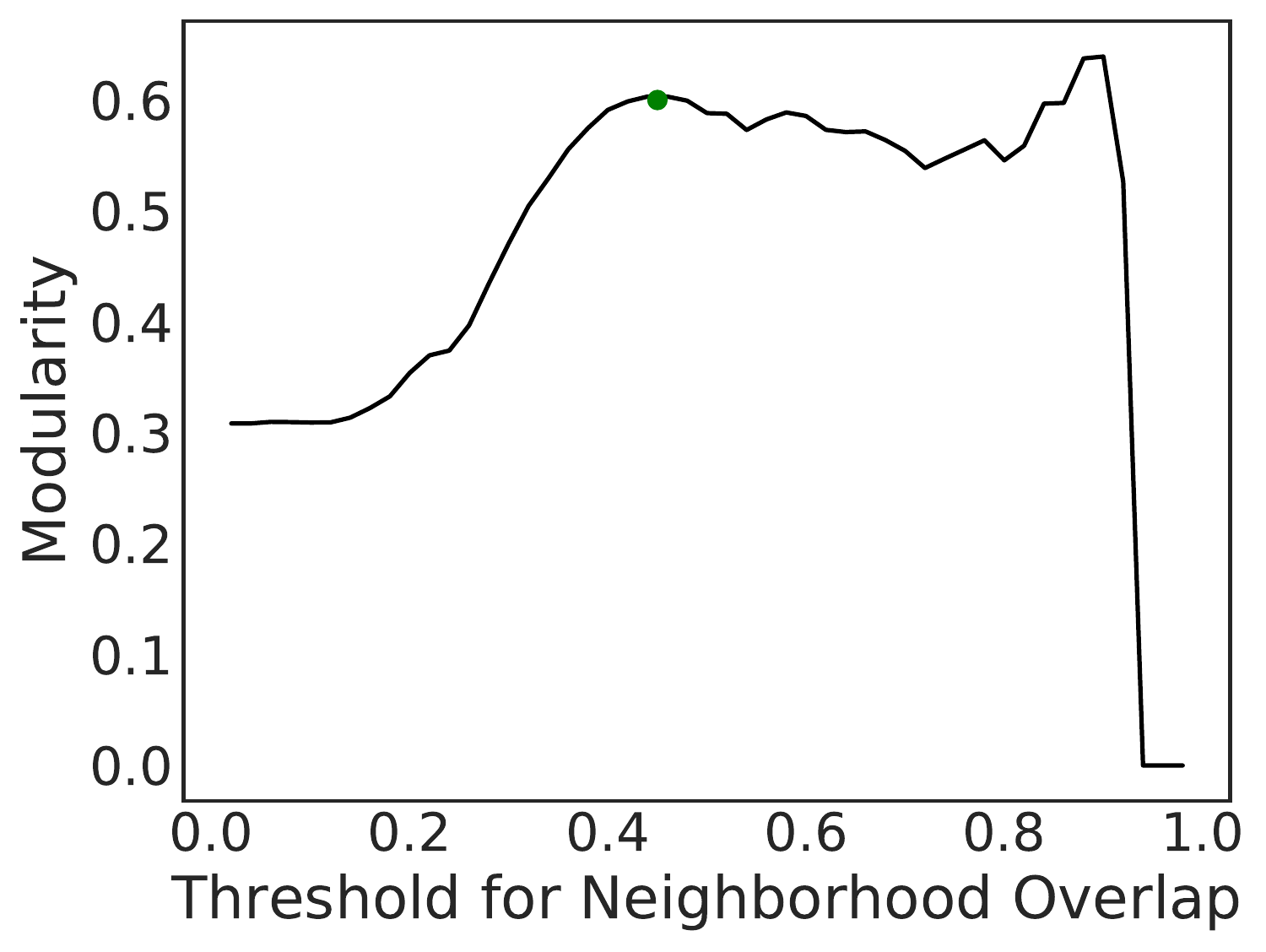} \label{Fig:NB_US}}
  \subfloat[Number of Members]{\includegraphics[width=4.85cm]{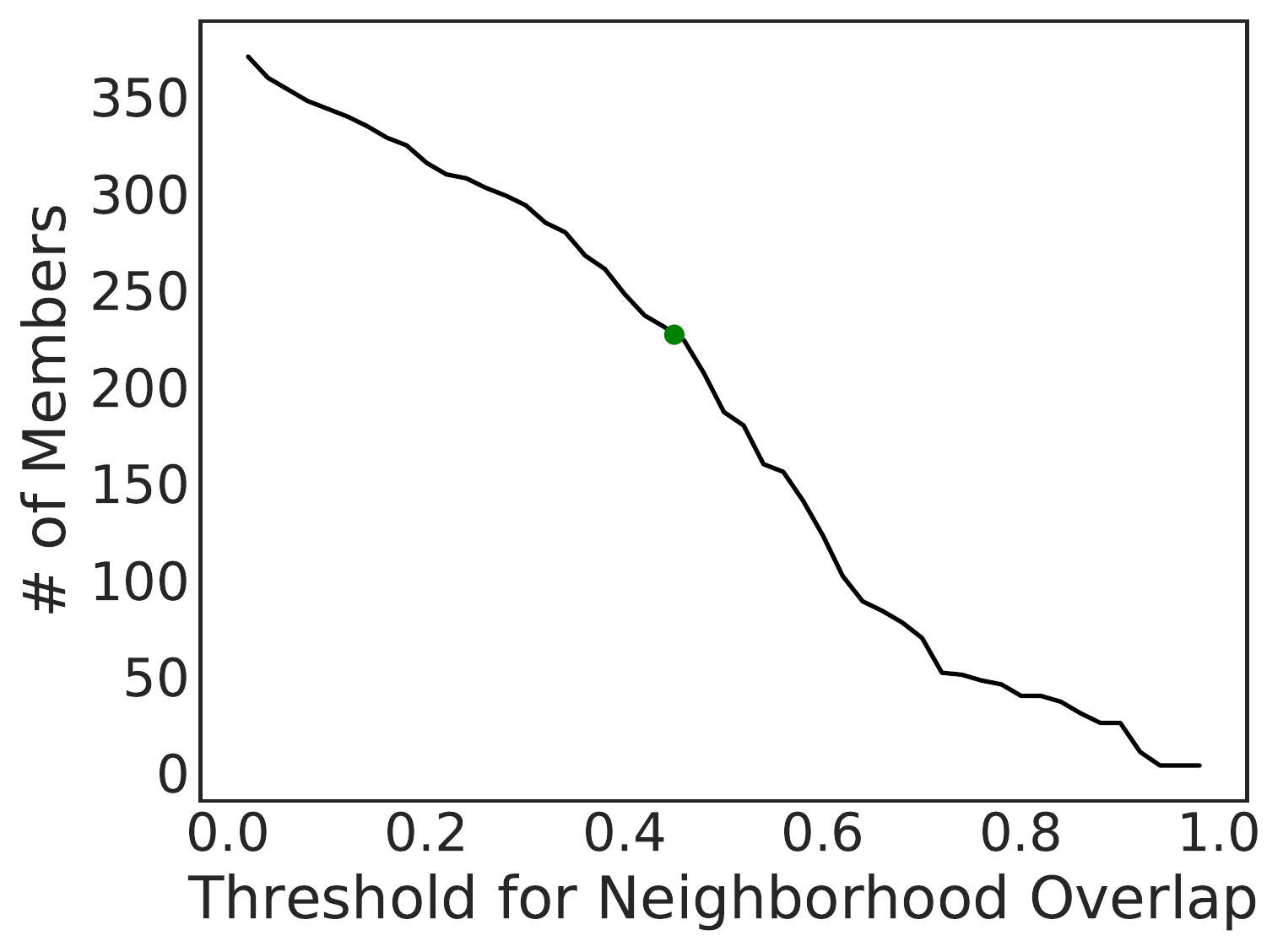} \label{Fig:NB_BR}}
  \vspace{-0.3cm}
  \caption{Impact of  Neighborhood Overlap Threshold for Brazil, 2017 (Selected threshold in green.).}
  \label{Fig:Tradeoff}
  \vspace{-0.8cm}
\end{figure}

Once again the selection of the best neighborhood overlap threshold was not clear as it involves a complex tradeoff: larger thresholds lead to more closely connected communities and higher \textit{modularity}, which is the goal,  but also produce more sparse graphs, resulting in a larger number of isolated nodes which are disregarded.  Thus, for each network, we selected a threshold that produced a good compromise between the two metrics. 
Figure \ref{Fig:Tradeoff}  shows an example of this trade-off for one specific year (2017) in Brazil, with the selected threshold value shown in green.
For   Brazil, the selected threshold fell between 0.40 and 0.55, while for the United States this range was from 0.1 to 0.28. We then  re-executed the Louvain algorithm to detect (polarized) communities in the new networks.


Table \ref{tab:RQ2} presents  the topological properties of the  networks as well as the structural and ideological properties of the identified polarized communities, for both Brazil and US. Focusing first on the Brazilian networks (top part of the table),  we see that  the number of nodes with strong ties decreases drastically (by up to 66\%) as compared to the networks analyzed in Section 4. This indicates the large presence of House members that, despite great similarity with other members, are not strongly tied (as defined above) to them, and thus do not belong to any polarized community. The number of connected components dropped for some years and increased for others, suggesting that some components in the first set of networks were composed of structurally weaker   communities or of multiple smaller communities. Network density, average shortest path length, and clustering coefficient also dropped, indicating more sparse networks, as expected.

\begin{table}[t]
\centering
\scriptsize
\caption{Characterization of Strongly Tied Networks and Polarized Communities}
\label{tab:RQ2}
\resizebox{.9\columnwidth}{!}{
\begin{tabular}{|c|c|c|c|c|c|c|c|c|c|c|c|}
\hline
\multicolumn{12}{|c|}{\textbf{Brazil}}                                                \\ \hline
\textbf{Year} & \textbf{\begin{tabular}[c]{@{}c@{}}\# of \\ Nodes\end{tabular}} & \textbf{\begin{tabular}[c]{@{}c@{}}\# of \\ Edges\end{tabular}} & \textbf{\begin{tabular}[c]{@{}c@{}}\# of \\ CC\end{tabular}} & \textbf{\begin{tabular}[c]{@{}c@{}}Avg.\\ SPL\end{tabular}} & \multicolumn{1}{c|}{\textbf{\begin{tabular}[c]{@{}c@{}}Avg.\\ Degree\end{tabular}}} & \textbf{\begin{tabular}[c]{@{}c@{}}Avg. \\ Clustering\end{tabular}} & \textbf{Density} & \textbf{\begin{tabular}[c]{@{}c@{}}\# of \\ Comm.\end{tabular}} & \textbf{Mod.} & \textbf{\begin{tabular}[c]{@{}c@{}}Avg.\\ PD\end{tabular}} & \textbf{\begin{tabular}[c]{@{}c@{}}SD\\ PD\end{tabular}} \\ \hline
2003          & 186                  & 1436                 & 1                                                            & 1.48                                                        & 15.44                                                          & 0.38                                                                & 0.08             & 4                                                               & 0.35          & 97.78\%                                                      & 0.86                                                     \\ \hline
2004          & 154                  & 866                  & 1                                                            & 1.52                                                        & 11.25                                                          & 0.33                                                                & 0.07             & 5                                                               & 0.36          & 97.11\%                                                      & 0.57                                                     \\ \hline
2005          & 119                  & 1210                 & 2                                                            & 1.19                                                        & 20.34                                                          & 0.59                                                                & 0.17             & 4                                                               & 0.37          & 95.40\%                                                      & 0.93                                                     \\ \hline
2006          & 136                  & 590                  & 10                                                           & 1.37                                                        & 8.68                                                           & 0.52                                                                & 0.06             & 12                                                              & 0.57          & 96.62\%                                                      & 2.16                                                     \\ \hline
2007          & 175                  & 977                  & 3                                                            & 1.68                                                        & 11.17                                                          & 0.32                                                                & 0.06             & 6                                                               & 0.44          & 97.31\%                                                      & 1.36                                                     \\ \hline
2008          & 216                  & 1019                 & 2                                                            & 1.94                                                        & 9.44                                                           & 0.23                                                                & 0.04             & 5                                                               & 0.42          & 97.11\%                                                      & 0.46                                                     \\ \hline
2009          & 209                  & 1217                 & 1                                                            & 1.30                                                        & 11.65                                                          & 0.41                                                                & 0.05             & 5                                                               & 0.56          & 94.57\%                                                      & 1.67                                                     \\ \hline
2010          & 225                  & 726                  & 6                                                            & 1.45                                                        & 6.45                                                           & 0.22                                                                & 0.02             & 11                                                              & 0.51          & 94.31\%                                                      & 1.80                                                     \\ \hline
2011          & 250                  & 1891                 & 1                                                            & 1.78                                                        & 15.13                                                          & 0.31                                                                & 0.06             & 4                                                               & 0.40          & 96.56\%                                                      & 0.86                                                     \\ \hline
2012          & 145                  & 1151                 & 3                                                            & 1.84                                                        & 29.82                                                          & 0.48                                                                & 0.11             & 6                                                               & 0.37          & 94.42\%                                                      & 1.98                                                     \\ \hline
2013          & 318                  & 4437                 & 5                                                            & 1.77                                                        & 27.91                                                          & 0.58                                                                & 0.08             & 9                                                               & 0.47          & 91.30\%                                                      & 2.17                                                     \\ \hline
2014          & 287                  & 1672                 & 3                                                            & 1.37                                                        & 11.65                                                          & 0.41                                                                & 0.04             & 5                                                               & 0.63          & 94.04\%                                                      & 1.28                                                     \\ \hline
2015          & 372                  & 6290                 & 6                                                            & 1.41                                                        & 33.82                                                          & 0.64                                                                & 0.09             & 9                                                               & 0.64          & 93.93\%                                                      & 1.70                                                     \\ \hline
2016          & 269                  & 1726                 & 3                                                            & 1.43                                                        & 12.83                                                          & 0.44                                                                & 0.04             & 8                                                               & 0.63          & 95.08\%                                                      & 1.21                                                     \\ \hline
2017          & 227                  & 1631                 & 5                                                            & 1.58                                                        & 14.37                                                          & 0.44                                                                & 0.06             & 6                                                               & 0.60          & 95.25\%                                                      & 2.01                                                     \\ \hline \hline  \hline 
\multicolumn{12}{|c|}{\textbf{United States}}                                                \\ \hline
\textbf{Year} & \textbf{\begin{tabular}[c]{@{}c@{}}\# of \\ Nodes\end{tabular}} & \textbf{\begin{tabular}[c]{@{}c@{}}\# of \\ Edges\end{tabular}} & \textbf{\begin{tabular}[c]{@{}c@{}}\# of \\ CC\end{tabular}} & \textbf{\begin{tabular}[c]{@{}c@{}}Avg.\\ SPL\end{tabular}} & \multicolumn{1}{c|}{\textbf{\begin{tabular}[c]{@{}c@{}}Avg.\\ Degree\end{tabular}}} & \textbf{\begin{tabular}[c]{@{}c@{}}Avg. \\ Clustering\end{tabular}} & \textbf{Density} & \textbf{\begin{tabular}[c]{@{}c@{}}\# of \\ Comm.\end{tabular}} & \textbf{Mod.} & \textbf{\begin{tabular}[c]{@{}c@{}}Avg.\\ PD\end{tabular}} & \textbf{\begin{tabular}[c]{@{}c@{}}SD\\ PD\end{tabular}} \\ \hline
2003          & 431                                                             & 41872                                                           & 2                                                            & 1.11                                                        & 194.30                                                                              & 0.95                                                                & 0.45             & 2                                                               & 0.47          & 93.60\%                                                      & 1.03                                                     \\ \hline
2004          & 426                                                             & 40741                                                           & 2                                                            & 1.12                                                        & 191.27                                                                              & 0.95                                                                & 0.45             & 2                                                               & 0.48          & 92.97\%                                                      & 0.55                                                     \\ \hline
2005          & 431                                                             & 41886                                                           & 2                                                            & 1.11                                                        & 194.37                                                                              & 0.95                                                                & 0.45             & 2                                                               & 0.47          & 92.60\%                                                      & 0.79                                                     \\ \hline
2006          & 426                                                             & 41073                                                           & 2                                                            & 1.10                                                        & 192.83                                                                              & 0.95                                                                & 0.45             & 2                                                               & 0.48          & 91.45\%                                                      & 0.33                                                     \\ \hline
2007          & 414                                                             & 38462                                                           & 2                                                            & 1.12                                                        & 185.81                                                                              & 0.94                                                                & 0.44             & 2                                                               & 0.42          & 91.55\%                                                      & 3.78                                                     \\ \hline
2008          & 423                                                             & 40708                                                           & 2                                                            & 1.11                                                        & 192.47                                                                              & 0.95                                                                & 0.45             & 2                                                               & 0.43          & 95.49\%                                                      & 1.93                                                     \\ \hline
2009          & 428                                                             & 41690                                                           & 2                                                            & 1.15                                                        & 194.81                                                                              & 0.94                                                                & 0.45             & 2                                                               & 0.40          & 93.89\%                                                      & 2.45                                                     \\ \hline
2010          & 418                                                             & 39958                                                           & 2                                                            & 1.13                                                        & 191.19                                                                              & 0.95                                                                & 0.45             & 3                                                               & 0.43          & 94.86\%                                                      & 1.97                                                     \\ \hline
2011          & 422                                                             & 41112                                                           & 2                                                            & 1.15                                                        & 194.84                                                                              & 0.97                                                                & 0.46             & 3                                                               & 0.45          & 90.01\%                                                      & 3.16                                                     \\ \hline
2012          & 413                                                             & 40529                                                           & 2                                                            & 1.07                                                        & 196.27                                                                              & 0.97                                                                & 0.47             & 3                                                               & 0.44          & 91.70\%                                                      & 2.17                                                     \\ \hline
2013          & 421                                                             & 40910                                                           & 2                                                            & 1.10                                                        & 194.35                                                                              & 0.96                                                                & 0.46             & 2                                                               & 0.46          & 93.32\%                                                      & 0.94                                                     \\ \hline
2014          & 417                                                             & 40717                                                           & 2                                                            & 1.08                                                        & 195.29                                                                              & 0.96                                                                & 0.46             & 2                                                               & 0.48          & 94.40\%                                                     & 0.38                                                     \\ \hline
2015          & 424                                                             & 41759                                                           & 2                                                            & 1.08                                                        & 196.98                                                                              & 0.95                                                                & 0.46             & 2                                                               & 0.47          & 94.53\%                                                      & 1.41                                                     \\ \hline
2016          & 418                                                             & 40890                                                           & 2                                                            & 1.08                                                        & 195.65                                                                              & 0.96                                                                & 0.46             & 3                                                               & 0.46          & 95.67\%                                                      & 0.80                                                     \\ \hline
2017          & 421                                                             & 40923                                                           & 2                                                            & 1.08                                                        & 194.41                                                                              & 0.95                                                                & 0.46             & 2                                                               & 0.48          & 95.37\%                                                      & 0.11                                                     \\ \hline
\end{tabular}
\vspace{-0.6cm}}
\end{table}

The number  of polarized communities somewhat differs from the number of communities obtained when all (strong and weak) ties are considered, increasing in most years. This 
suggests that some  ideological communities identified in Section 4 may be indeed formed by multiple more closely connected subgroups.  Yet, those numbers  are still smaller than the number of parties in each year (Table \ref{tab:Vote_Data}). Moreover, compared to the ideological communities first analyzed,  the polarized communities  are stronger both structurally and ideologically, as expressed by larger values of \textit{modularity} and average party discipline.

For the US case,  the numbers in Table \ref{tab:RQ2}  are very similar  to those in Table \ref{tab:RQ1}. Less than 2\% of the nodes have only weak ties and were removed from the networks in all years.
Thus,  almost all members have strong ties to each other, building ideological communities that are, in general, very polarized.

In sum, despite the fragmented party system, polarization can be observed in Brazil, to some degree, in a number of smaller strongly tied communities. In the US, on the other hand, almost all members and communities are  very polarized.


\section{Temporal Analysis}

We finally turn to RQ3 and investigate how the polarized communities evolve over time. To that end, we compute two complementary metrics, namely \textit{persistence} and normalized mutual information \cite{Vinh:2010, Wei:2015}, for each pair of consecutive years. We define  the {\it persistence} from year $x$ to  $x$+$1$ as the fraction of all members of polarized communities in $x$ who remained in some polarized community in $x$+$1$. A  \textit{persistence} equal to 100\% implies that all members of polarized communities in $x$ remained in some polarized community in $x+1$. Yet,   the membership of individual communities may have changed as members may have moved to  different communities.  To assess the extent of change in community membership over consecutive years, we compute the normalized mutual information (NMI) over the communities, taking only members who persisted over the two years. 

NMI is based on Shannon entropy of information theory \cite{Shannon:2001}. Given two sets of partitions $X$ and $Y$, defining community assignments for nodes, the mutual information   of $X$ and $Y$ can be thought as the informational ``overlap" between $X$ and $Y$, or how much we learn about $Y$ from $X$ (and about $X$ from $Y$). Let $P(x)$ be the probability that a node picked at random is assigned to community $x$, $P(x,y)$ be the probability that a node picked at random is assigned to both $x$ in $X$ and $y$ in $Y$. Also, let $H(X)$ be the Shannon entropy for $X$ defined as $H(X)=-\sum_x P(x) log P(x)$.  The NMI of $X$ and $Y$ is defined as:
\vspace{-0.2cm}
\begin{equation}
NMI(X,Y)=\frac{\sum_x \sum_y P(x,y) log \frac{P(x,y)}{P(x)P(y)}}{\sqrt{H(X)H(Y)}}
\end{equation}
\vspace{-0.2cm} 

NMI ranges from 0 (all members changed their communities) to 1 (all members remained in the same communities). 

Table \ref{tab:temporal_results} shows \textit{persistence} (\textit{Pers}) and NMI results for all pairs of consecutive years and  both countries. For Brazil (BR), the values of \textit{persistence} varied over the years, ranging from 46\% to 80\%. Thus,  a significant number of new nodes join polarized communities every year.  Indeed, in most years, roughly half of the members of polarized communities are newcomers.  Moreover, the values of NMI are small, especially in the earlier years, reflecting great change also in terms of nodes switching communities. This is consistent with a period of less clear distinction  between the communities and weaker polarization, as discussed in the previous sections. Since 2012, the values of NMI fall around 0.6, reflecting greater stability in community membership.  
For the US, on the other hand, both \textit{persistence} and NMI are very large, approaching the maximum (1). Almost all members persist in their polarized communities over the years.

\begin{table}[t]
\centering
\scriptsize
\caption{Temporal Evolution of Polarized Ideological Communities.}
\label{tab:temporal_results}
\begin{tabular}{|c|c|c|c|c|c|c|c|c|c|c|c|c|}
\hline
\multicolumn{2}{|c|}{\textbf{\begin{tabular}[c]{@{}c@{}}Sequential\\ Years\end{tabular}}} & \textbf{\begin{tabular}[c]{@{}c@{}}2003\\ 2004\end{tabular}} & \textbf{\begin{tabular}[c]{@{}c@{}}2004\\ 2005\end{tabular}} & \textbf{\begin{tabular}[c]{@{}c@{}}2005\\ 2006\end{tabular}} & \textbf{\begin{tabular}[c]{@{}c@{}}2007\\ 2008\end{tabular}} & \textbf{\begin{tabular}[c]{@{}c@{}}2008\\ 2009\end{tabular}} & \textbf{\begin{tabular}[c]{@{}c@{}}2009\\ 2010\end{tabular}} & \textbf{\begin{tabular}[c]{@{}c@{}}2011\\ 2012\end{tabular}} & \textbf{\begin{tabular}[c]{@{}c@{}}2012\\ 2013\end{tabular}} & \textbf{\begin{tabular}[c]{@{}c@{}}2013\\ 2014\end{tabular}} & \textbf{\begin{tabular}[c]{@{}c@{}}2015\\ 2016\end{tabular}} & \textbf{\begin{tabular}[c]{@{}c@{}}2016\\ 2017\end{tabular}} \\ \hline
\multirow{2}{*}{\textbf{BR}}                       & \textbf{Pers.}                       & 58.24\%                                                      & 46.30\%                                                      & 53.04\%                                                      & 68.26\%                                                      & 63.80\%                                                      & 61.38\%                                                      & 80.08\%                                                      & 67.87\%                                                      & 61.23\%                                                      & 57.85\%                                                      & 57.47\%                                                      \\ \cline{2-13} 
                                                   & \textbf{NMI}                         & 0.14                                                         & 0.16                                                         & 0.20                                                         & 0.22                                                         & 0.18                                                         & 0.26                                                         & 0.14                                                         & 0.59                                                         & 0.56                                                         & 0.65                                                         & 0.58                                                         \\ \hline
\multirow{2}{*}{\textbf{US}}                       & \textbf{Pers.}                       & 98.13\%                                                      & 90.80\%                                                      & 98.36\%                                                      & 97.57\%                                                      & 86.74\%                                                      & 96.24\%                                                      & 96.18\%                                                      & 96.76\%                                                      & 97.85\%                                                      & 97.63\%                                                      & 86.26\%                                                      \\ \cline{2-13} 
                                                   & \textbf{NMI}                         & 0.97                                                         & 0.97                                                         & 1.00                                                         & 1.00                                                         & 1.00                                                         & 0.94                                                         & 0.96                                                         & 0.80                                                         & 1.00                                                         & 0.97                                                         & 0.98                                                         \\ \hline
\end{tabular}
\vspace{-0.6cm}
\end{table}

\begin{figure}[t]
  \centering
  \subfloat[\textit{Brazil}]{\includegraphics[width=3.8cm]{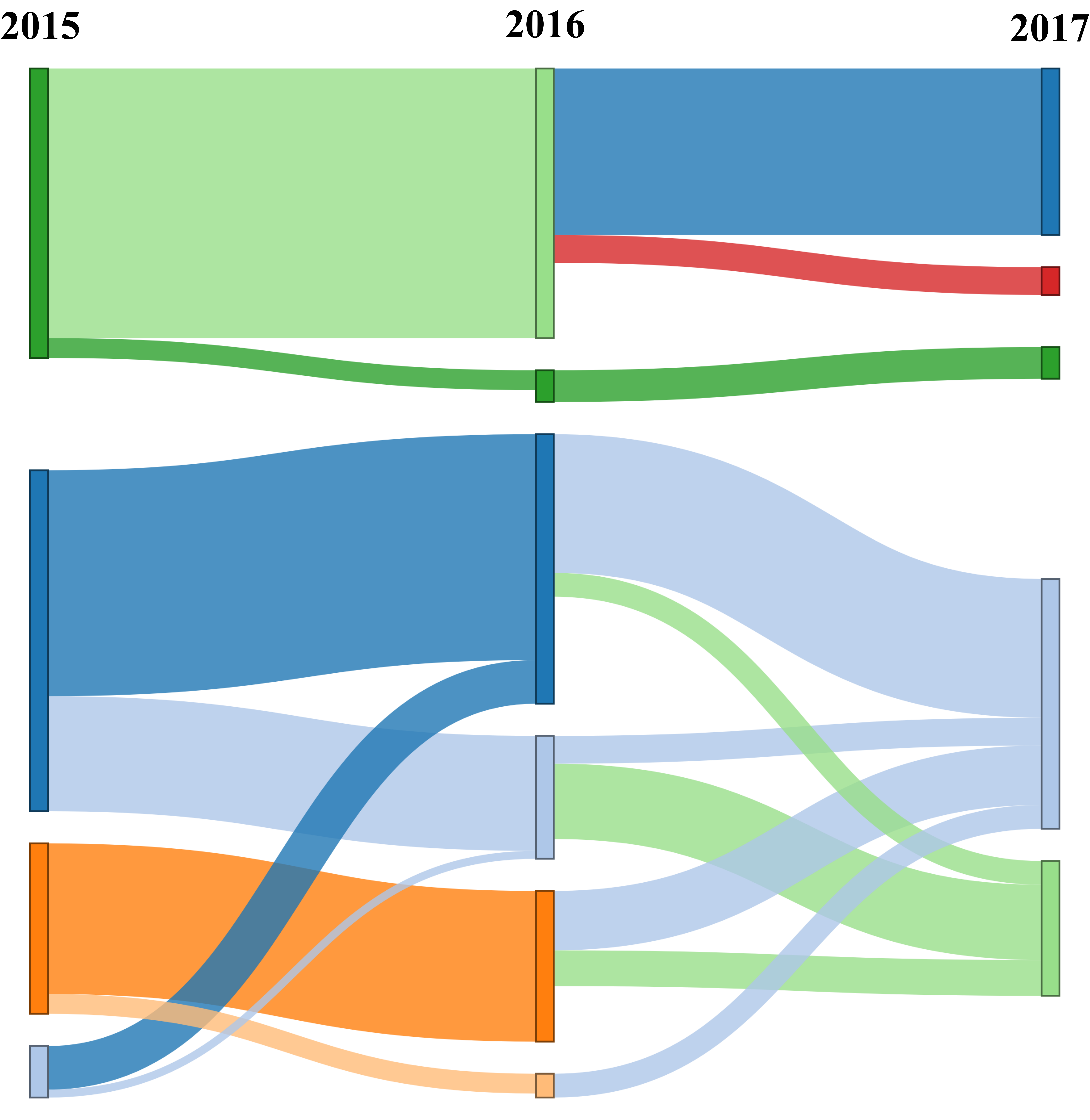} \label{Fig:Sankey_BR}}
  \subfloat[\textit{United States}]{\includegraphics[width=3.8cm]{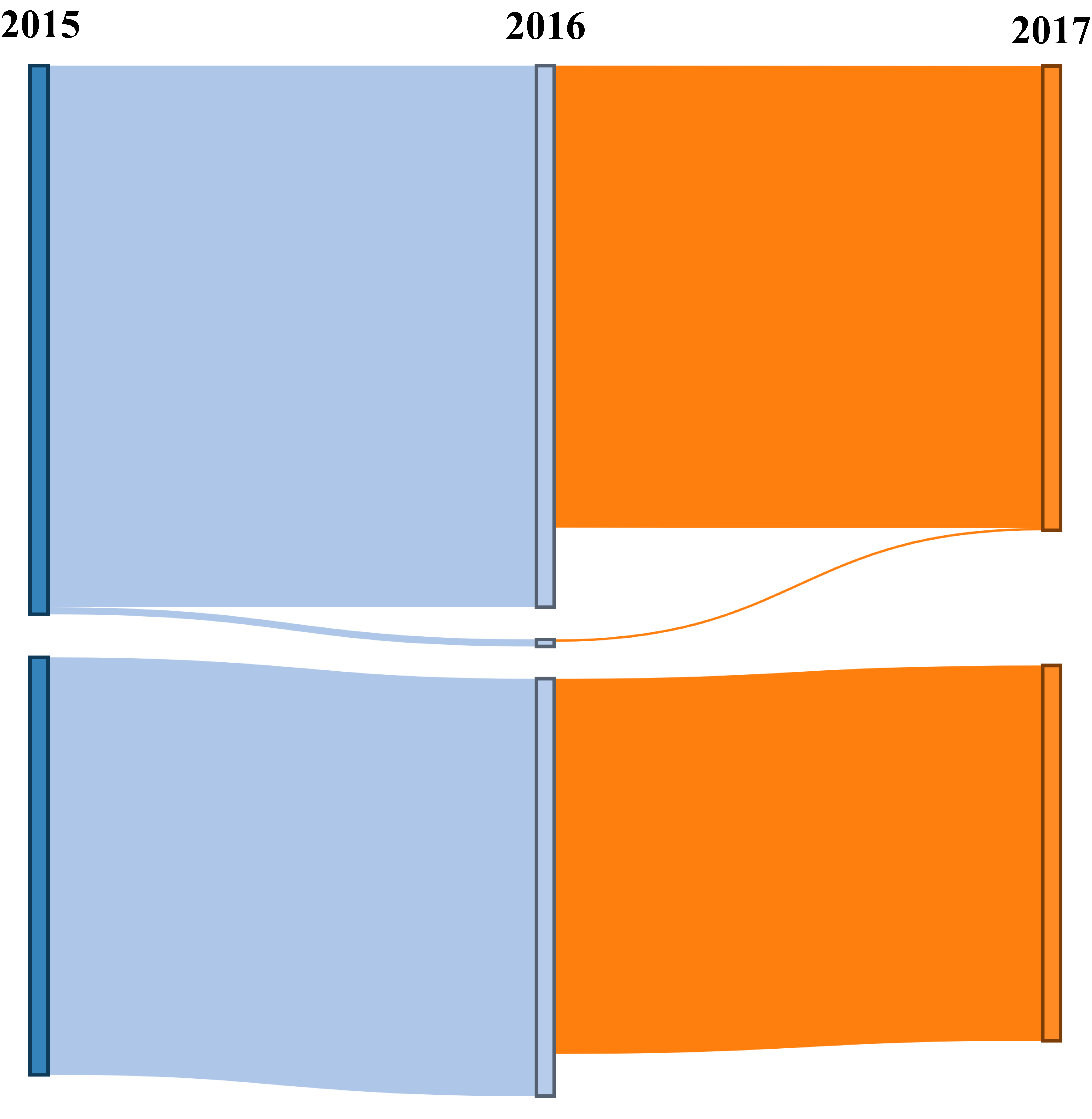} \label{Fig:Sankey_US}}
  \vspace{-0.3cm}
  \caption{  Dynamics of Polarized Communities over 2015-2017.}
  \label{Fig:Sankey}
  \vspace{-0.6cm}
\end{figure}

A visualization of some of these results is shown in Figure \ref{Fig:Sankey} which presents the flow of nodes across polarized communities over the years of 2015 to 2017 in Brazil and in the US.  Each vertical line represents a community, and its length   represents the number of members belonging to that community who persisted in some polarized community in the following year. Thus, communities for which all members did not persist in any polarized community in the following year are not represented in the figure.  Recall that, according to Table \ref{tab:RQ2}, the number of polarized communities in Brazil in 2015, 2016 and 2017 was 9, 8 and 6, respectively.  A cross-analysis of these results with Figure \ref{Fig:Sankey_BR} indicates that members of only 4 out of 9 polarized communities in 2015 persisted polarized in the following year. Moreover, two polarized communities in 2016 were composed of only  newcomers and both communities disappeared in 2017 (as they do not appear in the figure). Similarly, one polarized community in 2017 was composed of only newcomers.  The figure also shows a great amount of switching, merging and splitting across communities over the years.  Figure \ref{Fig:Sankey_US}, on the other hand, illustrates the greater stability of community membership in the US.



\section{Conclusions and Future Work}

We have proposed a methodology to analyze the formation and evolution of ideological and polarized communities in party systems, applying it to two strikingly different political contexts, namely Brazil and the US. Our analyses showed that the large number of political parties in Brazil can be reduced to only a few ideological communities, maintaining their original ideological properties, that is well disciplined communities, with a certain degree of redundancy.  These communities have distinguished themselves both structurally and ideologically in the recent years, a reflection of the transformation that Brazilian politics has been experiencing since 2012.  For the US, the country's strong and non-fragmented party system leads to the identification of ideological communities in the two main parties throughout the analyzed period.  However, there are still some highly similar links crossing the community boundaries. Moreover, for some years, a third community emerged, without however affecting the  strong discipline, ideology and community structure of the American party system.



We then took a step further and focused on polarized communities by considering only tightly connected groups of nodes. We found that in Brazil, despite the party fragmentation and the existence of some degree of similarity even across the identified ideological communities,  it is still possible to find a subset of members that organize themselves into strongly polarized ideological communities. However, these communities are highly dynamic, changing a large portion of their membership over consecutive years.  In the US, on the other hand, most ideological communities identified are indeed highly polarized and their membership remain mostly unchanged over the years.



As future work, we intend to further analyze ideological communities in our datasets by characterizing members in terms of their centrality as well as proposing new metrics of tie strength for this particular domain. We also intend to extend our study to other party systems. 

\vspace{-0.4cm}
\subsubsection*{Acknowledgments}
This work was partially supported by the FAPEMIG-PRONEX-MASWeb project --
Models, Algorithms and Systems for the Web, process number APQ-01400-14, as well as by the National Institute of Science and Technology for the Web (INWEB), CNPq and FAPEMIG.

\bibliographystyle{splncs03}
\bibliography{references.bib}
\end{document}